\documentclass[conference]{IEEEtran}

\IEEEoverridecommandlockouts

\usepackage{amsmath,amssymb,amsfonts,amsthm,dsfont} 
\usepackage{algorithm,algorithmicx,listings}  
\usepackage[noend]{algpseudocode}			  
\algnewcommand{\LineComment}[1]{\State \(\%\) #1 } 
\algrenewcommand{\algorithmiccomment}[1]{\hfill \(\%\) #1}
\usepackage{hyperref}					  

\usepackage{graphicx,color}

\usepackage{enumitem, ulem}	 
\usepackage{gensymb}		

\usepackage{extarrows}

\def\argmin{\mathop{\arg\,\min}\limits}	%

\DeclareMathOperator{\tr}{tr}

\renewcommand{\IEEEproofindentspace}{1\parindent}

\newtheorem{thm}{Theorem}

\newtheorem{definition}{Definition}
\newtheorem{assumption}{Assumption}

\newtheorem*{problem*}{Problem}

\newtheorem{lemma}{Lemma}

\begin{document}
\title{Information Acquisition with Sensing Robots: Algorithms and Error Bounds}

\author{Nikolay Atanasov, Jerome Le Ny, Kostas Daniilidis, and George J. Pappas
\thanks{This work was supported by ONR-MURI HUNT grant N00014-08-1-0696 and by TerraSwarm, one of six centers of STARnet, a Semiconductor Research Corporation program sponsored by MARCO and DARPA.}
\thanks{N. Atanasov and G. Pappas are with the Department of Electrical and Systems Engineering, University of Pennsylvania, Philadelphia, PA 19104, {\tt\small\{atanasov, pappasg\}@seas.upenn.edu}.}%
\thanks{J. Le Ny is with the Department of Electrical Engineering, Ecole Polytechnique de Montreal, ON, Canada, {\tt\small jerome.le-ny@polymtl.ca}}%
\thanks{K. Daniilidis is with the Department of Computer and Information Science, University of Pennsylvania, Philadelphia, PA 19104, {\tt\small kostas@cis.upenn.edu}}%
\thanks{This work has been submitted to the IEEE for possible publication. Copyright may be transferred without notice, after which this version may no longer be accessible.}%
}
\maketitle

\begin{abstract}
Utilizing the capabilities of configurable sensing systems requires addressing difficult information gathering problems. Near-optimal approaches exist for sensing systems without internal states. However, when it comes to optimizing the trajectories of mobile sensors the solutions are often greedy and rarely provide performance guarantees. Notably, under linear Gaussian assumptions, the problem becomes deterministic and can be solved \textit{off-line}. Approaches based on submodularity have been applied by ignoring the sensor dynamics and \textit{greedily} selecting informative locations in the environment. This paper presents a \textit{non-greedy} algorithm with \textit{suboptimality guarantees}, which does not rely on submodularity and takes the sensor dynamics into account. Our method performs provably better than the widely used greedy one. Coupled with linearization and model predictive control, it can be used to generate adaptive policies for mobile sensors with non-linear sensing models. Applications in gas concentration mapping and target tracking are presented.
\end{abstract}

\section{Introduction}
\label{sec:intro}
Miniaturization, wireless communication, and sensor technology have advanced remarkably in recent years. Autonomous vehicles instrumented with various sensors and interconnected by a network are deployed in large numbers, leading to configurable networked sensing systems. In order to utilize their capabilities, some important information gathering problems such as environmental monitoring (temperature, humidity, air quality, etc.) \cite{Choi_PhD09}, \cite{Leonard_IEEE09}, \cite{Hernandez_ICRA13}, surveillance and reconnaissance \cite{Ribsky_ICAA00}, \cite{Hilal_PhD13}, search and rescue \cite{Kumar_PerComp04}, active perception and active SLAM \cite{Roy_ICRA05}, \cite{karasev12_visual_learning}, \cite{Atanasov_ICRA13} need to be addressed.

\textit{Sensor management} \cite{Hero_Sensors11} offers a formal methodology to control the degrees of freedom of sensing systems in order to improve the information acquisition process. Much research in the field has been focused on sensors without internal states by studying the problem of \textit{sensor scheduling}, also referred to as \textit{sensor selection}. Efficient non-myopic approaches have been proposed \cite{ZAHV_Automatica09}, \cite{Boyd_TSP09}, \cite{William_PhD07}. However, when it comes to mobile sensors, which posses internal states, the results are limited. The main complication is that the evolution of the states depends on the management decisions and affects the measurements in the long run. Concretely, whereas in radar management a sensor can switch instantaneously between targets \cite{Krishnamurthy_TSP01}, a feasible and informative path needs to be designed for a sensing robot \cite{Singh_JAIR09}. Due to this complication, most approaches for mobile sensor management are myopic \cite{Grocholsky_PhD02}, \cite{Chung_ICRA06} or use short planning horizons \cite{Kreucher_PhD05}, \cite{Huber_PhD09}. However, it is precisely the presence of an internal state that makes multi-step optimization important. The behavior of the observed phenomenon needs to be predicted at an early stage to facilitate effective control of the mobile sensor.

A key insight is that under linear Gaussian assumptions the problem can be formulated as deterministic optimal control \cite{Jerome_CDC09}. As a result, informative sensing paths can be planned \textit{off-line}. Some of the successful approaches rely on a \textit{submodular} function to quantify the informativeness of the sensor paths \cite{Singh_JAIR09}, \cite{Singh_IJCAI09}. The sensing locations are partitioned into independent clusters. Submodularity is used to greedily select informative locations within clusters. An orienteering problem is solved to choose the sequence of clusters to visit. The drawback is that within clusters the movement of the sensor is restricted to a graph, essentially ignoring the sensor dynamics. As a result, the cluster sizes cannot be increased much without affecting the quality of the solution. We address this limitation by considering the sensor dynamics and planning non-greedily.

The contribution of this paper is an approach for discrete-time dynamic sensors with linear Gaussian sensing models to track a target with linear dynamics. We present an approximate non-myopic algorithm to decrease the complexity of obtaining an optimal policy, while providing strong performance guarantees. The generated suboptimal policy performs provably better than the widely used greedy one. Our work can be considered a search-based method for planning in information space. Related work in this area includes \cite{vanderhookTwoRobotTR}, \cite{Jerome_CDC09}. Sampling-based methods have been proposed as well \cite{LanSchwager_ICRA13}, \cite{Hollinger_RSS13}. Just as in traditional planning, they are able to find feasible solutions quickly but provide no finite-time guarantees on optimality. Approaches which do not make linear Gaussian assumptions and use non-linear filters exist as well \cite{Charrow_RSS13}, \cite{Hoffmann_TAC10}, \cite{LaValle_SensFilt12}. They can handle general sensing and target models but are computationally demanding and have no performance guarantees.

The rest of the paper is organized as follows. In Section \ref{sec:prob_form} we formulate the information acquisition problem precisely. In Section \ref{sec:sep_prin} we prove a separation principle, which allows computing the optimal sensor trajectory off-line via forward value iteration. We discuss how to reduce the exponential complexity and develop an approximate algorithm with suboptimality guarantees in Section \ref{sec:solution}. In Section \ref{sec:applications} we show how to generate adaptive policies for non-linear mobile sensors. Applications in gas distribution mapping and target tracking are presented. All proofs are provided in the Appendix.

\section{Problem Formulation}
\label{sec:prob_form}
\subsection{Information acquisition with sensing robots}
Consider a sensor mounted on a robot or vehicle, whose dynamics are governed by the following \textit{sensor motion model}:
\begin{equation} \label{eq:sensor_motion_model}
  x_{t+1} = f(x_t,u_t),
\end{equation}
where $x \in \mathcal{X} \cong \mathbb{R}^{n_x}$ is the sensor state, $\mathcal{X}$ is an $n_x$-dimensional state space with metric $d_\mathcal{X}$, $u \in \mathcal{U}$ is the control input, and $\mathcal{U}$ is a \textit{finite} space of admissible controls. The state of the vehicle is assumed known.

The task of the sensor is to track the evolution of the state of a target, whose dynamics are governed by a linear \textit{target motion model}:
\begin{equation} \label{eq:target_motion_model}
  y_{t+1} = Ay_t + w_t, \qquad w_t \sim \mathcal{N}(0,W),
\end{equation}
where $y \in \mathbb{R}^{n_y}$ is the target state, $A \in \mathbb{R}^{n_y \times n_y}$, and $w_t$ is a white Gaussian noise with covariance $\mathbb{E}w_t w_{\tau}^T = W \delta(t-\tau)$. The minimum eigenvalue of $W$ is denoted by $\uline{\lambda}_W$.

The operation of the sensor is governed by the following \textit{sensor observation model}:
\begin{equation} \label{eq:obs_model}
  z_t = H(x_t) y_t + v_t(x_t), \qquad v_t(x_t) \sim \mathcal{N}(0,V(x_t)),
\end{equation}
where $z_t \in \mathbb{R}^{n_z}$ is the measurement signal, $H(x_t) \in \mathbb{R}^{n_z \times n_y}$, and $v_t(x_t)$ is a sensor-state-dependent Gaussian noise, whose values at any pair of times are independent. The measurement noise is independent of the target noise $w_t$ as well. Note that the observation model is \textit{linear in the target state} but might depend nonlinearly on the sensor state, which typically reflects the fact that the sensor has a limited field of view. The information available to the sensor at time $t$ to compute the control $u_t$ is summarized in the information state:
\[
\mathcal{I}_0 := z_0 \quad \mathcal{I}_t := (z_{0:t}, u_{0:(t-1)}) \in \bigl(\mathbb{R}^{n_z}\bigr)^{t+1} \times \mathcal{U}^{t}, \; t>0.
\]

\begin{problem*}[Active Information Acquisition]
Given an initial sensor pose $x_0 \in \mathcal{X}$, a prior distribution of the target state $y_1$, and a planning horizon $T$, the task of the sensor is to choose a sequence of functions $\mu_0:\mathbb{R}^{n_z} \rightarrow \mathcal{U}$, $\mu_t: \mathbb{R}^{(t+1) n_z} \times \mathcal{U}^t \rightarrow \mathcal{U}$ for $t = 1,\ldots,T-1$, which maximize the mutual information between the final target state $y_{T+1}$ and the measurement set $z_{1:T}$:
\begin{alignat}{2}
\max_{\mu_0,\ldots,\mu_{T-1}} \; & \mathbb{I}(y_{T+1};z_{1:T}\mid x_{1:T}) && \label{eq:prob1}\\
\text{s.t.} \; \quad &x_{t+1} = f(x_t, \mu_t(\mathcal{I}_t)), \quad &&t = 0,\ldots,T-1, \notag\\
&y_{t+1} = Ay_t + w_t, \quad &&t = 1,\ldots,T, \notag\\
&z_t = H(x_t) y_t + v_t(x_t), \quad &&t = 1,\ldots,T. \notag
\end{alignat}
\end{problem*}

\textit{Remark:} To simplify notation, we work with autonomous models but all results hold for time-varying $f_t$, $A_t$, $W_t$, and $H_t$ with $\uline{\lambda}_W > 0$ such that $\uline{\lambda}_W I_{n_y} \preceq W_t$ for all $t \in [1,T]$.

Problem (\ref{eq:prob1}) subsumes the static case where sensors do not have internal states. In particular, if $f(x,u) := u$ and we think of the control $u$ as choosing a subset of sensors to be activated, then (\ref{eq:prob1}) becomes a sensor scheduling problem. Similarly, if the space of sensor configurations, $\mathcal{X}$, is restricted to a graph and revisiting is not allowed, we get the problem in \cite{Singh_JAIR09}. Numerous information gathering tasks can be cast in the form of (\ref{eq:prob1}). For example, environmental monitoring \cite{Jerome_CDC09}, active target tracking (Sec. \ref{sec:applications}, \cite{Charrow_RSS13}), and active mapping with gas (Sec. \ref{sec:applications}), stereo \cite{Freundlich_ICRA13}, laser, or any other sensor, whose operation is captured reasonably by a linearized model. 

To motivate the discussion consider the methane emission monitoring problem, which was addressed by the Best Service Robotics paper \cite{Hernandez_ICRA13} at ICRA 2013. The task is to estimate the gas concentration in a landfill using a remote methane leak detector (RMLD) based on tunable diode laser absorption spectroscopy. The RMLD sensor is mounted on a robotic platform, Gasbot, which follows an exploration path \textit{pre-specified by hand}. In this paper, we would like to automatically generate the most informative path for the Gasbot.

In detail, suppose that state of the Gasbot consists of its 2D position $(x^1,x^2) \in \mathbb{R}^2$ and the orientation of the RMLD sensor $\theta \in SO(2)$ so that $x := (x^1,x^2,\theta)$. At each time period, the Gasbot can move on a grid and choose the orientation of the sensor in $30\degree$ increments: $\Theta := \{-\pi, -5\pi/6,\ldots,5\pi/6\}$,
\[
\mathcal{U} := \biggl\{\begin{pmatrix}dx & dy & d\theta\end{pmatrix}^T \bigg\vert \begin{pmatrix}dx\\dy\end{pmatrix} \in \{0,\pm e_1, \pm e_2\}, d\theta \in \Theta\biggr\},
\]
where $e_1$ and $e_2$ are the standard basis vectors in $\mathbb{R}^2$. The sensor motion model is:
\begin{align*}
  x_{t+1} = \begin{pmatrix}x_{t+1}^1 & x_{t+1}^2 & \theta_{t+1} \end{pmatrix} = \begin{pmatrix}x_t^1 + dx & x_t^2 +dy & d\theta\end{pmatrix}.
\end{align*}
Given a time horizon $T$, we would like to choose a control policy for the Gasbot in order to obtain a good map $y_{T+1} \in \mathbb{R}^{n_y}$ of the gas concentration in the landfill. The $i$th component, $y_{T+1}^i$, represents the estimate of the concentration in parts per million (ppm) in the $i$th cell of the map. We assume a static methane field so that $A = I_{n_y}, W = 0$. It was experimentally verified in \cite{Hernandez_ICRA13} that a good sensor observation model is:
\[
z_t = H(x_t)y_t + v_t=\sum_{i=1}^{n_y}l_i(x_t) y_t^i + v_t, \quad v_t \sim \mathcal{N}(0,V),
\]
where the $i$th component of $H(x_t) \in \mathbb{R}^{1\times n_y}$ is the distance $l_i(x_t)$ traveled by sensor laser beam in cell $y_t^i$ for the given sensor pose $x_t$. Solving problem (\ref{eq:prob1}) will provide an automatic way to control the Gasbot in order to obtain an accurate map of the methane concentration.

\subsection{Notation}
The sets of $n \times n$ symmetric positive definite and semidefinite matrices are denoted $S^n_{++}$ and $S^n_{+}$ respectively. Given a control sequence $\sigma = u_{t},\ldots,u_{T-1} \in \mathcal{U}^{T-t}$ at time $t$, the corresponding sensor trajectory is $\pi(\sigma) := x_{t+1},\ldots,x_T \in \mathcal{X}^{T-t}$. Also, let $\sigma_i := u_{t+i}$, $\pi_{i+1} := x_{t+i+1}$, and $\pi(i+1) := x_{t+i+1},\ldots, x_T \in \mathcal{X}^{T-t-i}$ for $i = 0,\ldots,(T-t-1)$.

\section{A Separation Principle}
\label{sec:sep_prin}
\subsection{Open-loop control is optimal} 
Active Information Acquisition (\ref{eq:prob1}) is a stochastic optimal control problem and in general for such problems adaptive (closed-loop) control policies have a significant advantage over non-adaptive (open-loop) ones. However, due to the linearity of the observation model (\ref{eq:obs_model}) in the target state and the Gaussian-noise assumptions, it can be shown that (\ref{eq:prob1}) reduces to a deterministic optimal control problem, for which open-loop policies are optimal.
\begin{thm}
\label{thm:separation}
If the prior distribution of $y_1$ is Gaussian with covariance $\Sigma_0 \in S_+^{n_y}$, there exists an \textit{open-loop} control sequence $\sigma \!=\! u_0,\ldots,u_{T-1} \in \mathcal{U}^T\!$, which is optimal in (\ref{eq:prob1}). Moreover, (\ref{eq:prob1}) is equivalent to the following deterministic optimal control problem:
\begin{alignat}{2}
\min_{\sigma \in \mathcal{U}^T} \; &\log\det(\Sigma_T) &&\label{eq:prob2}\\
\text{s.t.} \quad &x_{t+1} = f(x_t, \sigma_t), \quad &&t = 0,\ldots,T-1, \notag\\
&\Sigma_{t+1} = \rho_{x_{t+1}}(\Sigma_t), \quad &&t = 0,\ldots,T-1, \notag
\end{alignat}
where $\rho_x(\cdot) := \rho^p(\rho_x^e(\cdot))$ is the Kalman filter Riccati map:
\begin{flalign}
\textbf{Update: } \rho_x^e(\Sigma) &:= (\Sigma^{-1} + M(x))^{-1} = C_x(\Sigma)\Sigma &\notag\\
&\phantom{:}=C_x(\Sigma)\Sigma C_x^T(\Sigma) + K_x(\Sigma)V(x)K_x^T(\Sigma)&\notag\\
M(x) &:= H(x)^T V(x)^{-1} H(x)& \label{eq:sensor_info_mat}\\
C_x(\Sigma) &:= I - K_x(\Sigma)H(x) = (I + \Sigma M_x)^{-1}&\notag\\
K_x(\Sigma) &:= \Sigma H(x)^TR_x^{-1}(\Sigma)&\notag\\
R_x(\Sigma) &:= H(x) \Sigma H(x)^T + V(x)&\notag\\
\textbf{Predict: } \rho^p(\Sigma) &:= A\Sigma A^T + W& \notag
\end{flalign}
and $M(\cdot) \in \mathbb{R}^{n_y \times n_y}$ is called the \textit{sensor information matrix}.
\end{thm}
\textit{Remark}: Our solution to (\ref{eq:prob2}) is applicable to any \textit{monotone concave} cost function but for clarity we use $\log\det(\cdot)$.
\textit{Remark}: the conditional mean of $y_{t+1}$ given $z_{0:t}$ can also be computed recursively via the Kalman filter:
\begin{flalign*}
\textbf{Update: } \xi^e(y) &:= y + K_x(\Sigma)(z - H(x)y)&\\
\textbf{Predict: } \xi^p(y) &:= Ay&
\end{flalign*}
but does not play a role in the optimization (\ref{eq:prob2}).

\subsection{Forward value iteration}
Separation principles like Thm. \ref{thm:separation} have been exploited since the early work on sensor management \cite{Athans_Auto72}, \cite{Meier_TAC67}. The result is significant because the state space of a deterministic control problem is much smaller than the stochastic version (distributions over hidden variables are not needed) and an open-loop policy can be computed \textit{off-line}. As discussed in \cite{Jerome_CDC09}, the main bottleneck for computations is the large dimension of the state $(x_t,\Sigma_t)$ and it is beneficial to use forward value iteration (Alg. \ref{alg:val_iter}). The advantage is that the set of reachable covariance matrices is built progressively starting from the initial state. As a result, only the reachable subspace is considered instead of discretizing the whole space of covariances as required by backward value iteration.

\begin{algorithm}[htb]
\caption{Forward Value Iteration}
\label{alg:val_iter}
\begin{algorithmic}[1]
\footnotesize
\State $S_0 \gets \{(x_0,\Sigma_0)\}$, \quad $S_t \gets \emptyset$ for $t = 1,\ldots,T$
\For{$t=1:T$}
	\ForAll{$(x,\Sigma) \in S_{t-1}$}
		\ForAll{$u \in \mathcal{U}$}
			\State $x_t \gets f(x,u)$
			\State $S_t \gets S_t \cup \{(x_t,\rho_{x_t}(\Sigma))\}$ 
		\EndFor
	\EndFor
\EndFor
\State \textbf{return} $\min \; \{\log\det(\Sigma) \mid (x,\Sigma) \in S_T\}$
\end{algorithmic}
\end{algorithm}
The forward value iteration (FVI) is constructing a \textit{search tree}, $\mathcal{T}_t$, with nodes at stage $t \in [0,T]$ corresponding to the reachable states $(x_t,\Sigma_t)$. The algorithm starts from the initial node $(x_0,\Sigma_0)$ and uses the controls from $\mathcal{U}$ to obtain the set of nodes $S_1$ reachable at time $t=1$. In general, starting from node $(x_t,\Sigma_t)$, there is an edge for each control in $\mathcal{U}$ leading to a node $(x_{t+1}, \Sigma_{t+1})$ obtained from the dynamics in (\ref{eq:prob2}). Even though FVI provides a computational advantage to the backward version, the number of nodes in $\mathcal{T}_t$ corresponds to the number of sensor trajectories of length $t$ and grows exponentially. Alg. \ref{alg:val_iter} is guaranteed to find the optimal control sequence $\sigma^*$ but has exponential complexity, $\mathcal{O}(|U|^T)$. The other extreme is the greedy policy $\sigma^g$: 
\begin{equation}
\sigma_t^g \in \argmin_{u \in \mathcal{U}} \biggl(\log\det \bigl(\rho_{f(x_t,u)}(\Sigma_t)\bigr)\biggr), \; t \in [0,T-1], \label{eq:greedy_policy}
\end{equation}
which is computationally very efficient, $\mathcal{O}(|\mathcal{U}|T)$, but no guarantees exist for its performance. We can think of it as a modification of Alg. \ref{alg:val_iter}, which discards all but the best (lowest cost) node at level $t$ of the tree $\mathcal{T}_t$.

The goal of this paper is to provide an algorithm, with complexity lower than that of FVI and performance better than that of the greedy policy (\ref{eq:greedy_policy}). This can be achieved by discarding some but not all of the nodes at level $t$ of $\mathcal{T}_t$. Intuitively, if two sensor trajectories cross, i.e. there are two nodes at level $t$ of $\mathcal{T}_t$ with the same vehicle configuration $x$ but different covariances, and one is clearly less informative, i.e. its covariance is dominated by the other one, then it is not necessary to keep it. The following section will make this intuition precise.

\section{Reduced Value Iteration}
\label{sec:solution}
\subsection{Optimality-preserving reductions}
A notion of redundancy for the nodes in the search tree $\mathcal{T}$ is provided by the following definition.

\begin{definition}[Algebraic redundancy \cite{ZAHV_Automatica09}]
\label{def:alg_red}
Let $\{\Sigma_i\}_{i = 1}^K \subset S^n_+$ be a finite set. A matrix $\Sigma \in S^n_+$ is algebraically redundant with respect to $\{\Sigma_i\}$, if there exist nonnegative constants $\{\alpha_i\}_{i=1}^{K}$ such that:
\[
\textstyle{\sum_{i=1}^{K} \alpha_i = 1 \quad\text{and}\quad \Sigma \succeq \sum_{i=1}^{K} \alpha_i \Sigma_i}.
\]
\end{definition}

\textit{Remark:} Definition \ref{def:alg_red} is a simplified version of the one in \cite{ZAHV_Automatica09}. As shown in the proof of Theorem \ref{thm:opt_prune}, the simplification is possible because the objective function in (\ref{eq:prob2}) minimizes the estimation error only at the final stage.

The next theorem shows that when several sensor paths cross at time $t$, i.e. there are several nodes at level $t$ of $\mathcal{T}_t$ with the same vehicle state, algebraically redundant ones can be discarded without removing the optimal one.

\begin{thm}[Optimal reduction]
\label{thm:opt_prune}
For $t \in [1,T]$, let $(x,\Sigma) \in S_t$ be a node in level $t$ of the search tree $\mathcal{T}_t$. If there exist a set $\{x^i,\Sigma^i\} \subseteq S_t \setminus \{(x,\Sigma)\}$ such that $x = x^i, \; \forall i$ and $\Sigma$ is algebraically redundant with respect to $\{\Sigma^i\}$, then $(x,\Sigma)$ can be removed from $S_t$ without eliminating the optimal trajectory.
\end{thm}

\subsection{\texorpdfstring{$\epsilon$}{epsilon}-Suboptimal reductions}
At the expense of losing optimality, we can discard even more of the crossing trajectories by using a relaxed notion of algebraic redundancy.

\begin{definition}[$\epsilon$-Algebraic redundancy \cite{ZAHV_Automatica09}]
\label{def:eps_alg_red}
Let $\epsilon \geq 0$ and let $\{\Sigma_i\}_{i=1}^K \subset S^n_+$ be a finite set. A matrix $\Sigma \in S^n_+$ is $\epsilon$-algebraically redundant with respect to $\{\Sigma_i\}$, if there exist nonnegative constants $\{\alpha_i\}_{i=1}^{K}$ such that:
\[
\textstyle{\sum_{i=1}^{K} \alpha_i = 1 \quad\text{and}\quad \Sigma+\epsilon I_n \succeq \sum_{i=1}^{K} \alpha_i \Sigma_i}.
\]
\end{definition}

Let $\pi^* = x_1^*,\ldots,x_T^* \in \mathcal{X}^T$ be the optimal sensor trajectory in $\mathcal{T}_T$ with covariance sequence $\Sigma_1^*,\ldots,\Sigma_T^*$ and cost $V_T^* := \log\det(\Sigma_T^*)$. Denote the search tree at time $t$ with all $\epsilon$-algebraically redundant nodes removed by $\mathcal{T}_t^\epsilon$. Let $\pi^\epsilon = x_1^\epsilon,\ldots,x_T^\epsilon \in \mathcal{X}^T$ be the trajectory obtained by forward value iteration on the reduced tree $\mathcal{T}_T^\epsilon$ with a corresponding covariance sequence $\Sigma_1^\epsilon,\ldots,\Sigma_T^\epsilon$ and cost $V_T^\epsilon := \log\det(\Sigma_T^\epsilon)$. The following theorem provides an upper bound on the gap, $(V_T^\epsilon - V_T^*)$, between the performances of $\pi^\epsilon$ and $\pi^*$.

\begin{thm}[$\epsilon$-Suboptimal reduction]
\label{thm:eps_red}
Let $\beta_* < \infty$ be the peak estimation error of the optimal policy, $\Sigma_t^* \preceq \beta_* I_{n_y}$, for $t \in [1,T]$. Then,
\begin{equation}
\label{eq:eps_guarantee}
0 \leq V_T^\epsilon - V_T^* \leq \epsilon \Delta_T,
\end{equation}
where
\[
\Delta_T := \frac{n_y}{\uline{\lambda}_W}\biggr(1 + \frac{\beta_*^2}{\uline{\lambda}_W^2}\bigl(1 - \eta_*^{T-1}\bigr)\biggl), \quad \eta_* := \frac{\beta_*}{\beta_*+\uline{\lambda}_W}< 1.
\]
\end{thm}

\textit{Remark}: If the peak estimation error $\beta_*$ remains bounded as $T \to \infty$, i.e. the sensor performs well when using the optimal policy, then $\Delta_T \to \frac{n_y}{\uline{\lambda}_W}\biggr(1 + \frac{\beta_*^2}{\uline{\lambda}_W^2}\biggl)$. In other words, the suboptimality gap of $\pi^\epsilon$ is bounded regardless of the length $T$ of the planning horizon and grows linearly in $\epsilon$!

\subsection{\texorpdfstring{$(\epsilon,\delta)$}{(epsilon, delta)}-Suboptimal reductions}
If the motion of the sensor is restricted to a graph many of the planned trajectories will be crossing and the results from Thm. \ref{thm:eps_red} are very satisfactory. However, if the space of sensor configurations, $\mathcal{X}$, is continuous, depending on the sensor motion model (e.g. differential drive), it is possible that no two sensor states reachable at time $t$ are exactly equal. Then, the reductions from Thm. \ref{thm:eps_red} cannot be applied. To avoid this case, we can relax the notion of trajectory crossing at time $t$.

\begin{definition}[Trajectory $\delta$-Crossing]
\label{def:gamma_cross}
Trajectories $\pi^1, \pi^2 \in \mathcal{X}^T$ $\delta$-cross at time $t \in [1,T]$ if $d_\mathcal{X}(\pi_t^1,\pi_t^2) \leq \delta$ for $\delta \geq 0$.
\end{definition}

Instead of discarding $\epsilon$-algebraically redundant trajectories which cross, we can discard those which $\delta$-cross. Let $\mathcal{T}_t^{\epsilon, \delta}$ be the reduced tree at time $t$, with all $\epsilon$-algebraically redundant $\delta$-crossing nodes removed. Some continuity assumptions are necessary in order to provide suboptimality guarantees for searching within $\mathcal{T}_t^{\epsilon, \delta}$.

\begin{assumption}[\textbf{Motion Model Continuity}]
\label{ass:smm_cont}
The sensor motion model is Lipschitz continuous in $x$ with Lipschitz constant $L_f \geq 0$ for every fixed $u \in \mathcal{U}$:
\[
d_\mathcal{X}(f(x_1,u),f(x_2,u)) \leq L_f d_\mathcal{X}(x_1,x_2), \quad \forall x_1, x_2 \in \mathcal{X}.
\]
\end{assumption}

\begin{assumption}[\textbf{Observation Model Continuity}]
\label{ass:som_cont}
There exists a real constant $L_m \geq 0$ such that:
\[
\begin{aligned}
M(x_1) &\preceq \bigl(1 + L_m d_\mathcal{X}(x_1,x_2)\bigr) M(x_2)\\
M(x_2) &\preceq \bigl(1 + L_m d_\mathcal{X}(x_1,x_2)\bigr) M(x_1)
\end{aligned}, \quad \forall x_1, x_2 \in \mathcal{X},
\]
where $M(\cdot)$ is the sensor information matrix (\ref{eq:sensor_info_mat}).
\end{assumption}

Assumption \ref{ass:smm_cont} says that when two sensor configurations are close and the same sequence of controls is applied, then the resulting trajectories will remain close. Assumption \ref{ass:som_cont} says that sensing from similar configurations results in similar information gain. This gives the intuition that when two trajectories $\delta$-cross their future informativeness will be similar. We make this intuition precise in the next theorem. Let $\pi^{\epsilon,\delta} \in \mathcal{X}^T$ be the sensor trajectory obtained by searching the reduced tree $\mathcal{T}_T^{\epsilon,\delta}$ with corresponding cost $V_T^{\epsilon,\delta}$. The gap, $(V_T^{\epsilon, \delta} - V_T^*)$, between the performances of $\pi^{\epsilon,\delta}$ and $\pi^*$ is bounded as follows.

\begin{thm}[$(\epsilon,\delta)$-Suboptimal reduction]
\label{thm:eps_del_red}
Let $\beta_* < \infty$ be the peak estimation error of the optimal policy, $\Sigma_t^* \preceq \beta_* I_{n_y}$, for $t \in [1,T]$. Then,
\begin{equation}
\label{eq:eps_del_guarantee}
0 \leq V_T^{\epsilon, \delta} - V_T^* \leq (\zeta_T - 1)\biggl[V_T^* - \log\det(W)\biggr] + \epsilon \Delta_T,
\end{equation}
\begin{flalign*}
&\text{where} \qquad \zeta_t := \prod_{\tau=1}^{t-1} \biggl(1+ \sum_{s=1}^\tau L_f^s L_m \delta \biggr) \geq 1,&\\
&\Delta_T := \frac{n_y}{\uline{\lambda}_W}\biggr(1 + \frac{\beta_*}{\uline{\lambda}_W}\sum_{\tau=1}^{T-1} \frac{\zeta_T}{\zeta_\tau} \eta_*^{T-\tau}\biggl), \quad \eta_* := \frac{\beta_*}{\beta_*+\uline{\lambda}_W}< 1.&
\end{flalign*}
\end{thm}

Notice that Thm. \ref{thm:eps_red} is a special case of Thm. \ref{thm:eps_del_red} because if $\delta=0$, then $\zeta_t = 1, \forall t \in [1,T]$ and (\ref{eq:eps_del_guarantee}) reduces to (\ref{eq:eps_guarantee}). Then, the suboptimality gap remains bounded regardless of the time horizon and grows linearly with $\epsilon$. If $\delta>0$, then $\lim_{T\to\infty}\zeta_T = \infty$ and the suboptimality gap grows with $T$ and $\delta$. The bound is loose, however, because it uses a worst-case analysis. The worst case happens when a trajectory, which was discarded from $\mathcal{T}_T^{\epsilon,\delta}$, persistently obtains more information than a trajectory, which remains very close in space and is still in the search tree. Even then, if the optimal policy performs well, the term $V_T^* - \log\det(W)$ gets smaller as $\zeta_T$ increases and the suboptimality gap remains small.

\subsection{\texorpdfstring{$(\epsilon,\delta)$}{(epsilon, delta)}-Reduced value iteration}
The approaches for reducing the search tree, developed in the previous subsections, can be used to significantly decrease the complexity of the FVI algorithm (Alg. \ref{alg:val_iter}), while providing suboptimality guarantees (Thm. \ref{thm:eps_del_red}). The $(\epsilon,\delta)$-Reduced Value Iteration (RVI) is summarized in Alg. \ref{alg:red_val_iter}.
\begin{algorithm}[H]
\caption{$(\epsilon,\delta)$-Reduced Value Iteration}
\label{alg:red_val_iter}
\begin{algorithmic}[1]
\footnotesize
\State $S_0 \gets \{(x_0,\Sigma_0)\}$, \quad $S_t \gets \emptyset$ for $t = 1,\ldots,T$
\For{$t=1:T$}
	\ForAll{$(x,\Sigma) \in S_{t-1}$}
		\ForAll{$u \in \mathcal{U}$}
			\State $x_t \gets f(x,u)$
			\State $S_t \gets S_t \cup \{(x_t,\rho_{x_t}(\Sigma))\}$
		\EndFor
	\EndFor
	\State Sort $S_t$ in ascending order according to $\log\det(\cdot)$
	\State $S_t' \gets S_t[1]$ \Comment{State with lowest cost}
	\ForAll{$(x,\Sigma) \in S_t \setminus S_t[1]$}
	  \LineComment{Find all nodes in $S_t'$, which $\delta$-cross $x$:}
	  \State $Q \gets \{ \Sigma' \mid (x',\Sigma') \in S_t', \; d_\mathcal{X}(x,x') \leq \delta\}$
	  \If{isempty($Q$) \textbf{or} \textbf{not}( $\Sigma$ is $\epsilon$-alg. redundant wrt $Q$ )}
	    \State $S_t' \gets  S_t' \cup (x,\Sigma)$
	  \EndIf
	\EndFor
	\State $S_t \gets S_t'$
\EndFor
\State \textbf{return} $\min \; \{\log\det(\Sigma) \mid (x,\Sigma) \in S_T\}$
\end{algorithmic}
\end{algorithm}
The most computationally demanding operation is checking $\epsilon$-algebraic redundancy (Line 12). It is a feasibility problem for a linear matrix inequality (LMI) and off-the-shelf solvers exist \cite{MATLAB_RCT}. An appealing property of Alg. \ref{alg:red_val_iter} is that at stage $t$ at least the node with currently lowest cost is retained (Line 8). This guarantees that the control sequence obtained from RVI \textit{performs at least as well as} the greedy policy (\ref{eq:greedy_policy}).

\section{Applications}
\label{sec:applications}
\subsection{Gas distribution mapping and leak localization}
\begin{figure*}[ht!]
	\begin{center}
		\includegraphics[width=0.62\linewidth,height=0.3\linewidth,trim=0mm 0mm 0mm 2.5mm,clip]{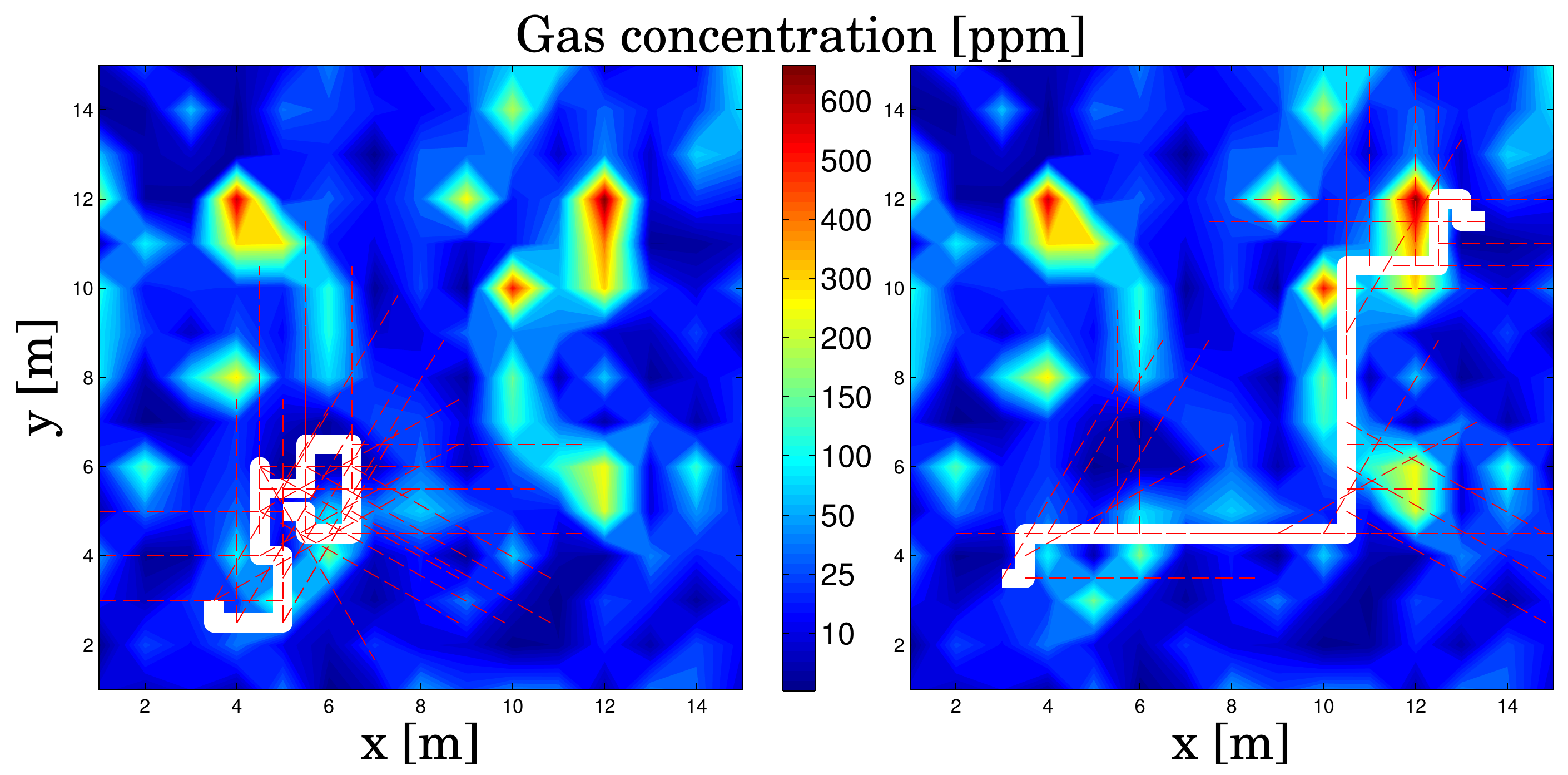}
		\includegraphics[width=0.34\linewidth]{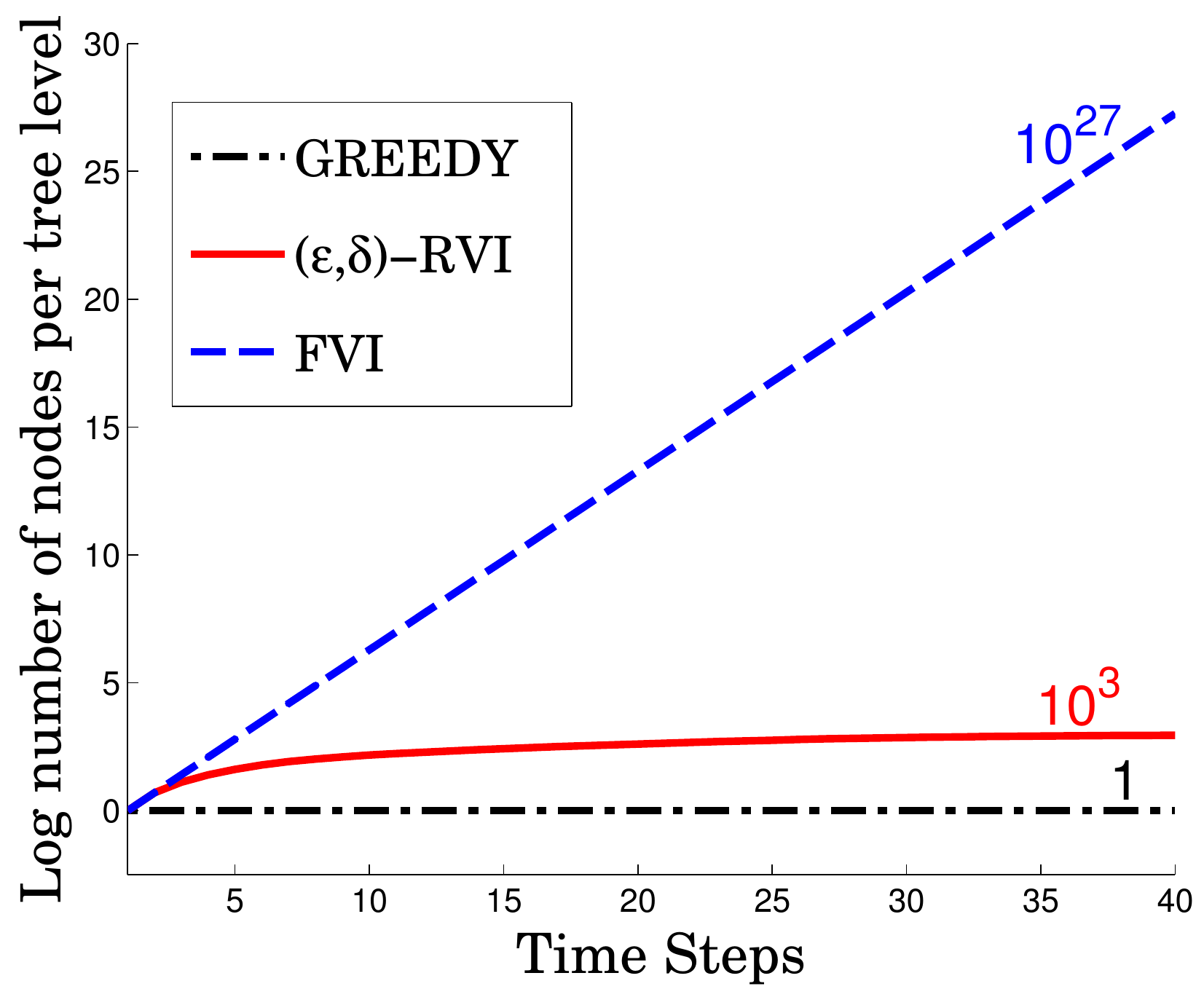}
	\end{center}
	\caption{Comparison of the sensor trajectories (white) obtained by the greedy algorithm (left) and the reduced value iteration (middle) with $\epsilon = \infty$ and $\delta=0$ after $40$ time steps. A typical realization of the methane field is shown. The red lines indicate the orientation of the gas sensor during the execution. On the right, the log number of nodes maintained in the search tree by the two approaches is compared to the complete tree maintained by forward value iteration.}
	\label{fig:gas_experiment}
\end{figure*}

In this subsection, we apply the $(\epsilon,\delta)$-RVI to the methane monitoring problem intorduced in Section \ref{sec:prob_form}. Since the movement of the Gasbot is restricted to a grid, the planned sensor paths will be crossing frequently and we can use $\delta = 0$. The dimension $n_y$ of the target is the number of cells in the gas concentration map and would typically be very large. Checking algebraic redundancy requires solving an $n_y$-dimensional LMI feasibility problem, which is computationally very demanding. To avoid this, we let $\epsilon = \infty$. This means that when several paths cross at time $t$, only the most informative one is kept in $\mathcal{T}_t^{\epsilon,\delta}$. Thus, the number of nodes in $\mathcal{T}_t^{\epsilon,\delta}$ remains bounded by the number of reachable sensor states. Trajectories of length $T = 40$ were planned using RVI and the greedy algorithm (GREEDY). The results (Fig. \ref{fig:gas_experiment}) reveal an important drawback of GREEDY. It remains trapped in a local region of relatively high variance and fails to see that there are more interesting regions which should be explored during the limited available time. Fig. \ref{fig:gas_experiment} also shows that the growth of the search tree is much slower for RVI compared to FVI, while the quality of the RVI solution is better than the greedy one.
 
\subsection{Target localization and tracking}
In a lot of applications, the linear assumptions (\ref{eq:target_motion_model}), (\ref{eq:obs_model}) are reasonable for the target motion model but very limiting for the sensor observation model. A lot more problems can fit in a framework with the following non-linear observation model:
\begin{equation}
z_t = h(x_t, y_t) + v(x_t, y_t), \; v(x_t, y_t) \sim \mathcal{N}(0, V(x_t, y_t)). \label{eq:nonlin_model}
\end{equation}
In this subsection, we show that our method can be coupled with linearization and model predictive control to generate an adaptive policy for a mobile sensor with the model in (\ref{eq:nonlin_model}).

\begin{figure*}[ht!]
	\begin{center}
		\includegraphics[width=0.24\linewidth,trim=0mm 0mm 0mm 7.5mm,clip]{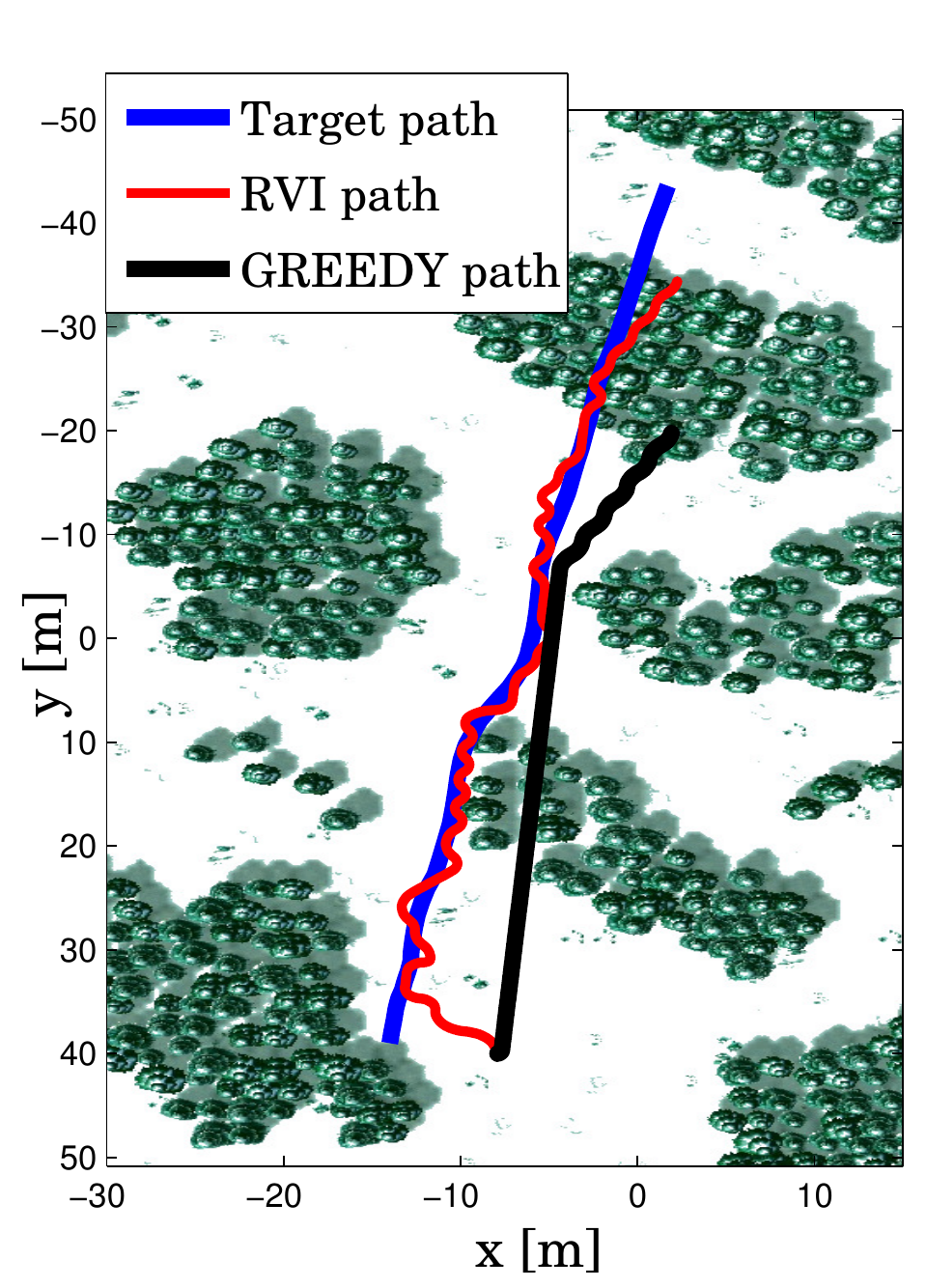}
		\includegraphics[width=0.36\linewidth]{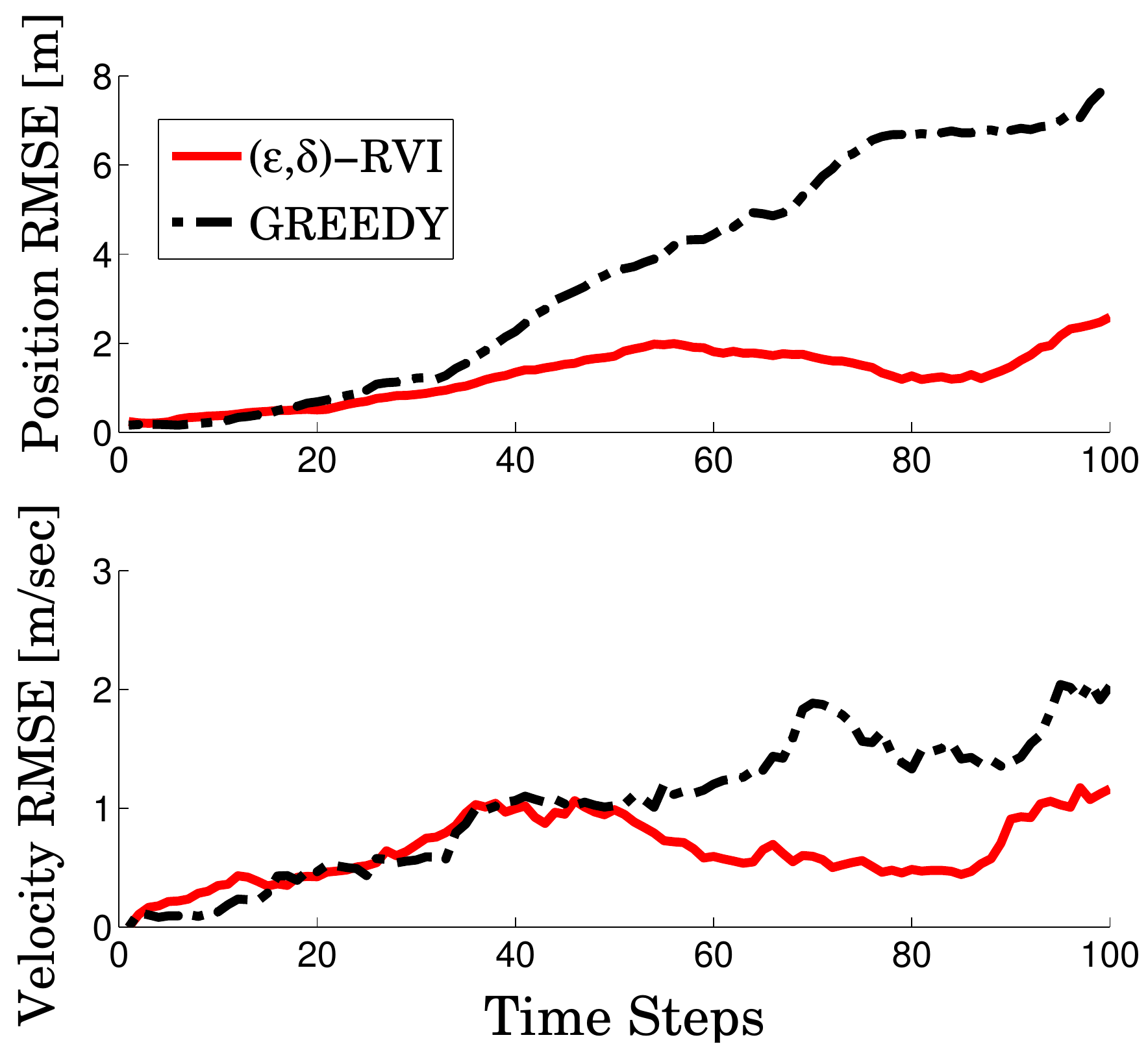}
		\includegraphics[width=0.36\linewidth]{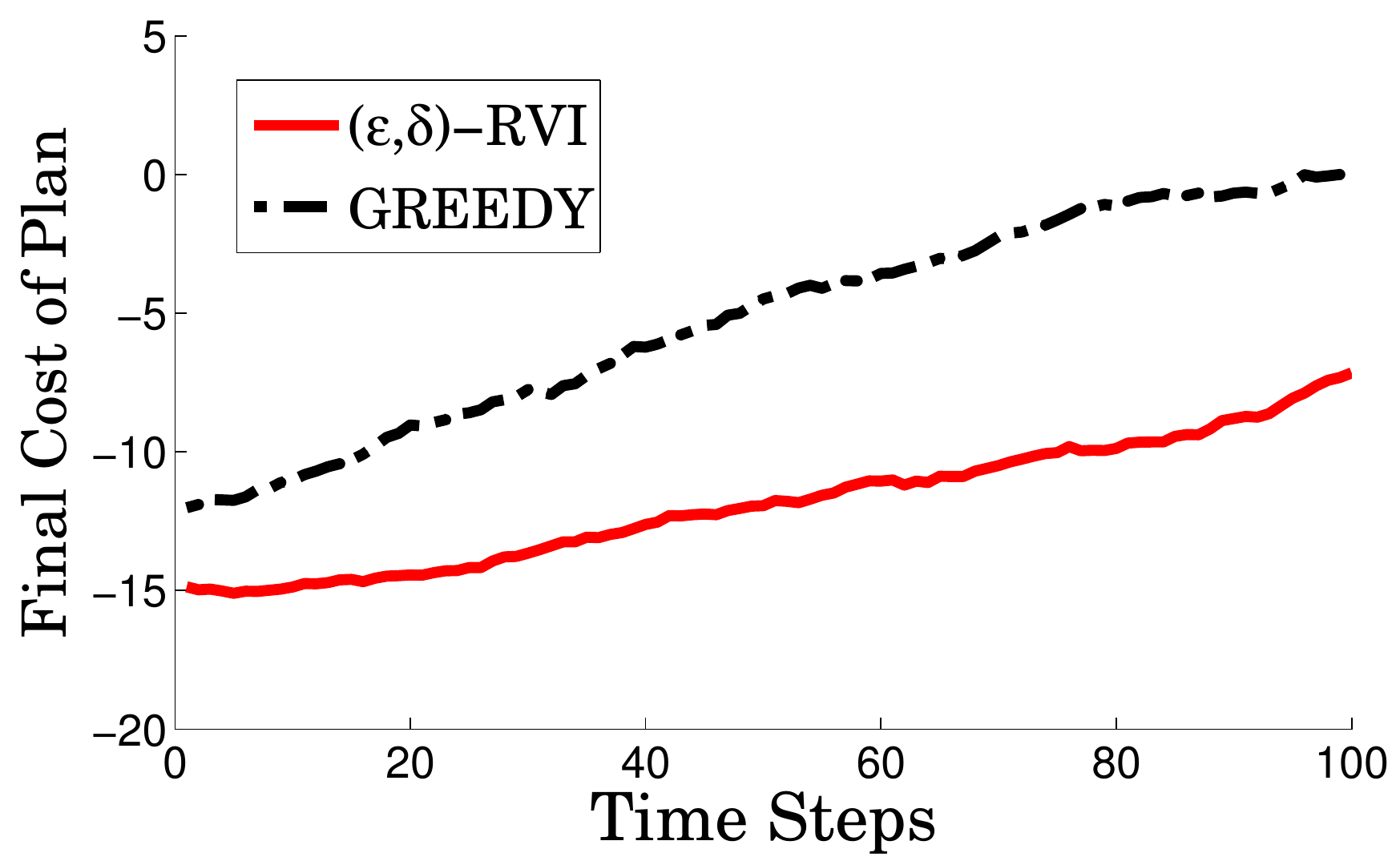}
	\end{center}
	\caption{Simulation results from 100 Monte-Carlo runs of the target tracking scenario. A typical realization is shown on the left. The average root-mean-square error (RMSE) of the estimated target position and velocity is shown in the middle. The $\log\det$ of the predicted target covariance is shown on the right.}
	\label{fig:dd_chase_experiment}
\end{figure*}

To motivate the discussion we consider a target tracking application, in which both the sensor motion and observation models are highly non-linear. Suppose that the sensor is mounted on a vehicle with differential-drive dynamics, which are discretized using a sampling period $\tau$ as follows:
\[
\begin{pmatrix}
x_{t+1}^1\\
x_{t+1}^2\\
\theta_{t+1}
\end{pmatrix} = \begin{pmatrix}
x_t^1\\
x_t^2\\
\theta_t
\end{pmatrix} + \begin{cases}
\begin{pmatrix}
\tau v\cos(\theta_t + \tau \omega/2)\\
\tau v\sin(\theta_t + \tau \omega/2)\\
\tau \omega
\end{pmatrix}, \; |\tau \omega| < 0.001\\
\begin{pmatrix}
\frac{v}{\omega}(\sin(\theta_t+\tau \omega) - \sin \theta_t)\\
\frac{v}{\omega}(\cos \theta_t - \cos(\theta_t + \tau \omega))\\
\tau \omega
\end{pmatrix}, \text{ else}.
\end{cases}
\]
The vehicle is controlled using the motion primitives $\mathcal{U} = \{(v,\omega) \mid v \in \{0, 1, 2, 3\} \text{ m/s}, \; \omega \in \{0,\pm \pi/2, \pm \pi\} \text{ rad/s}\}$. The task of the sensor is to track the position $(y^1, y^2) \in \mathbb{R}^2$ and velocity $(\dot{y}^1, \dot{y}^2) \in \mathbb{R}^2$ of another vehicle with discretized double integrator dynamics driven by Gaussian noise:
\begin{gather*}
y_{t+1} \!=\! \begin{bmatrix}
  I_2 & \tau I_2\\
  0 & I_2
\end{bmatrix} y_t \!+\! w_t, \;\; w_t \!\sim\! \mathcal{N}\biggl(\!0,q \begin{bmatrix}
  \tau^3/3 I_2 & \tau^2/2 I_2\\
  \tau^2/2 I_2 & \tau I_2
\end{bmatrix}\!\biggr),
\end{gather*}
where $y = [y^1, y^2, \dot{y}^1, \dot{y}^2]^T$ is the target state and $q$ is a scalar diffusion strength measured in $(\frac{m}{sec^2})^2 \frac{1}{Hz}$. The sensor takes noisy range and bearing measurments of the target's position:
\begin{equation}
h(x,y) \!=\! \begin{bmatrix}
r(x,y)\\
\alpha(x,y)
\end{bmatrix} \!:=\! \begin{bmatrix}
\sqrt{(y^1 - x^1)^2 + (y^2-x^2)^2}\\
\arctan\bigl((y^2-x^2)/(y^1 - x^1)\bigr)
\end{bmatrix}\label{eq:rb_model}
\end{equation}
The target needs to be tracked during a period $T_{max}$ in a wooded area (see Fig. \ref{fig:dd_chase_experiment}), which affects the covariance of the measurement noise. The noise in the range measurement grows linearly with the distance between the sensor and the target but trees along the way make the growth faster. The bearing measurement noise increases linearly with the speed of the sensor. Thus, good range measurements require that the sensor is close to the target and not blocked by trees, while good bearing measurements require that the sensor moves slowly. The sensor has a \textit{maximum range} of $15$ meters, after which the noise covariance is infinite. 

To exploit the independence of the covariance recursion of the Kalman filter from measurement values (Theorem \ref{thm:separation}) and to employ RVI, the observation model (\ref{eq:rb_model}) needs to be linearized about a predicted target trajectory during planning. Linearized about an arbitrary target state $y \neq x$, the model is:
\[
H(x,y) \!:=\! \nabla_y h(x,y) \!=\! \frac{1}{r(x,y)} \begin{bmatrix}
    (y^1-x^1) & (y^2-x^2)\\
    -\sin\alpha(x,y) & \cos\alpha(x,y)
\end{bmatrix}.
\]
The linearization depends on the predicted target trajectory, which in turn depends on the measurements obtained on-line. As a result, it is necessary to re-plan the sensor path on-line after every new measurement. In general, re-planning would be feasible only for a short horizon $T < T_{max}$ before the plan is needed. Alg. \ref{alg:lin_mpc} shows how to use the $(\epsilon,\delta)$-RVI and model predictive control to generate an adaptive policy.
\begin{algorithm}[H]
\caption{Model predictive control via the $(\epsilon,\delta)$-RVI}
\label{alg:lin_mpc}
\begin{algorithmic}[1]
\footnotesize
\State \textbf{Input}: $T_{max},x_0,\hat{y}_0,\hat{\Sigma}_0,f,\mathcal{U},H, V, A, W, T, \epsilon, \delta$
\For{$t=0:T_{max}$}
  \State Predict a target trajectory of length $T$: $\hat{y}_t,\ldots,\hat{y}_{t+T}$
  \State Linearize the observation model: $H_{\tau}(\cdot) \gets H(\cdot, \hat{y}_{t+\tau})$, $\tau = 0,..,T$
  \LineComment{\textit{Plan a sensor trajectory of length $T$}}
  \State $\sigma \gets (\epsilon,\delta)\text{-RVI}(x_t,\hat{y}_t,\hat{\Sigma}_t,\{H_{\tau}\}_{\tau=0}^T,V, f,\mathcal{U},A,W, T)$
  \State Move the sensor: $x_{t+1} \gets f(x_t,\sigma_1)$
  \State Get a new observation: $z_{t+1} \gets h(x_{t+1},y_{t+1}) + v_{t+1}(x_{t+1})$
  \State Update the target estimate: $(\hat{y}_{t+1},\hat{\Sigma}_{t+1}) \gets Filter(\hat{y}_t,\hat{\Sigma}_T,z_{t+1})$
\EndFor
\end{algorithmic}
\end{algorithm}
We emphasize that the linearized sensing models are utilized merely for determining the next configuration (Line 6), while the target state inference (Line 9) can still be performed with \textit{any non-linear filter}. Alg. \ref{alg:lin_mpc} was implemented with $T_{max} = 100$, $T=7$, $\epsilon = 0.1$, $\delta = 1$, $\tau = 0.5$, $q = 0.2$ and $100$ Monte-Carlo simulations were carried out. The tracking performance is compared to the greedy solution in Fig. \ref{fig:dd_chase_experiment}. We can see that the greedy policy goes in straight line to keep the speed low, i.e. the bearing noise small, but cannot predict in advance that the range noise will increase as the target goes further away. As a result, the greedy solution is likely to lose the target.

\section{Conclusion}
Under linear Gaussian assumptions, information acquisition with mobile sensors can be planned off-line. Most previous approaches are greedy or neglect the sensor dynamics and rarely provide performance guarantees. In this paper, we developed a non-greedy algorithm (RVI), which takes the sensor dynamics into account and has suboptimality guarantees. It provides parameters ($\epsilon, \delta$) to control the trade-off between complexity and optimality. Coupled with linearization and model predictive control, RVI can generate an adaptive policy for a non-linear mobile sensor. Our method can also be applied to a multi-sensor problem by stacking all motion and observation models into single centralized versions. Unfortunately, the complexity of this solution is exponential in the number of sensors. Future work will focus on a decentralized method for the multi-sensor, multi-target active information acquisition. In detail, we would like to address cooperative control, distributed target estimation, and noisy self-localization.

\appendices
\section*{Appendix A: Proof of Theorem \ref{thm:separation}}
\begin{lemma}
\label{lem:diff_ent_gauss}
Let $X\sim \mathcal{N}(\mu,\Sigma)$ be $n$-dimensional. Then, its differential entropy is $\mathbf{h}(X) = \bigl(n\log(2\pi e) + \log\det(\Sigma)\bigr)/2$.
\end{lemma}

\begin{IEEEproof}[Proof of Thm \ref{thm:separation}]
By definition of mutual information:
\[
\mathbb{I}(y_{T+1};z_{1:T}\mid x_{1:T}) = \mathbf{h}(y_{T+1} \mid x_{1:T}) - \mathbf{h}(y_{T+1} \mid z_{1:T}, x_{1:T}).
\]
Since $y_{T+1}$ is independent of the sensor path the first term is constant with respect to the optimization in (\ref{eq:prob1}). Due to linearity of the observation model (\ref{eq:obs_model}) in the target state, the distribution of $y_{t+1}$ given $z_{1:t}$ and $x_{1:t}$ remains Gaussian for $t\!>\!1$. Its covariance $\Sigma_t$ can be obtained from the Bayes filter, which due to the linear Gaussian assumptions is equivalent to the Kalman filter. Thus, $\Sigma_{t+1} = \rho_{x_t}(\Sigma_t)$ for $t = 0, \ldots,T-1$, which is \textit{independent} of the measurements $z_{1:t}$. By Lemma \ref{lem:diff_ent_gauss}:
\begin{align}
&\mathbf{h}(y_{T+1} \mid \!z_{1:T}, x_{1:T}) \!=\! \mathbb{E}_{\hat{z}_{1:T}} \mathbf{h}(y_{T+1} \mid z_{1:T} \!=\! \hat{z}_{1:T}, x_{1:T}) \label{eq:indep_entropy}\\
&\!= \frac{1}{2}\mathbb{E}_{\hat{z}_{1:T}}\!\bigl(\log(2\pi e)^{n_y} \!\!+\! \log|\Sigma_T|\bigr) = \frac{1}{2}\bigl(\log(2\pi e)^{n_y} \!\!+\! \log|\Sigma_T|\bigr). \notag
\end{align}
Let $\mu^* \!=\! \{\mu_0^*,..,\mu_{T-1}^*\}$ be optimal in (\ref{eq:prob1}) with cost $V^*$. Fix a realization $\hat{z}_{1:T}$ of the measurements and let $\sigma$ be the open-loop policy induced by $\mu^*$ given $\hat{z}_{1:T}$ with cost $V^\sigma$. From (\ref{eq:indep_entropy}), $V^*$ is independent of $\hat{z}_{1:T}$, hence $V^* = V^\sigma$ for any $\hat{z}_{1:T}$.
\end{IEEEproof}

\section*{Appendix B: Proof of Theorem \ref{thm:opt_prune}}
\begin{definition}[Operator monotone function]
Let $\Sigma_1, \Sigma_2 \in S^n_+$. A function $g: S^n_+ \rightarrow S^n_+$ is operator monotone if $\Sigma_1 \preceq \Sigma_2$ implies $g(\Sigma_1) \preceq g(\Sigma_2)$.
\end{definition}

\begin{definition}[Operator concave function]
Let $\{\Sigma_i \mid i = 1,\ldots,K\} \subset S^n_+$ be a finite set and let $\sum_{i=1}^K \alpha_i = 1$ for some real constants $\alpha_i \geq 0$. A function $g: S^n_+ \rightarrow S^n_+$ is operator concave if it satisfies $\sum_{i=1}^K \alpha_i g(\Sigma_i) \preceq g(\sum_{i=1}^K \alpha_i \Sigma_i)$.
\end{definition}

\begin{definition}[$t$-step Riccati map]
For $\pi \in \mathcal{X}^T,\;\Sigma \in S^n_+$ let
\[
  \phi_{\pi}^0(\Sigma) := \Sigma, \qquad \phi_{\pi}^t(\Sigma) := \rho_{\pi_t} \circ \ldots \circ \rho_{\pi_1}(\Sigma), \;\; t \in [1,T].
\]
\end{definition}

\begin{lemma}[\cite{ZAHV_Automatica09}]
\label{lem:ric_mon_con}
For any $t \in [0,T]$, the $t$-step Riccati map is operator monotone and operator concave.
\end{lemma}

\begin{IEEEproof}[Proof of Thm \ref{thm:opt_prune}]
Let $\sigma \in U^{T-t}$ be any admissible control sequence. Starting at $(x,\Sigma)$, by Lemma \ref{lem:ric_mon_con} and Definition \ref{def:alg_red}, there exist nonnegative constants $\{\alpha_i\}_{i=1}^K$ such that
\begin{align*}
\phi_{\pi(\sigma)}^{T-t}(\Sigma) \succeq \phi_{\pi(\sigma)}^{T-t}\biggl(\sum_{i=1}^{K} \alpha_i \Sigma^i\biggr) \succeq \sum_{i=1}^{K} \alpha_i \phi_{\pi(\sigma)}^{T-t}\bigl(\Sigma^i\bigr).
\end{align*}
Then, from monotonicity and concavity of $\log\det(\cdot)$:
\begin{align*}
\log&\det\biggl(\phi_{\pi(\sigma)}^{T-t}(\Sigma)\biggr) \geq \log\det \biggl(\sum_{i=1}^{K} \alpha_i \phi_{\pi(\sigma)}^{T-t}(\Sigma^i)\biggr)\\
&\geq \sum_{i=1}^{K} \alpha_i \log\det\biggl(\phi_{\pi(\sigma)}^{T-t}(\Sigma^i)\biggr) \geq \log\det\biggl(\phi_{\pi(\sigma)}^{T-t}(\Sigma^{i^*})\biggr).
\end{align*}
The last inequality holds because a convex combination of scalars is lower bounded by the smallest one $i^*$.
\end{IEEEproof}

\section*{Appendix C: Proof of Theorem \ref{thm:eps_del_red}}

\begin{lemma}[\cite{ZAHV_Automatica09}]
\label{lem:ric_dir_der}
For any $\pi \in \mathcal{X}^T$, $Q \in S^n_+$, and $\epsilon \geq 0$ the directional derivative of the $t$-step Riccati map is:
\begin{align*}
g_{\pi}^t(\Sigma;Q) &:= \frac{d}{d\epsilon} \phi_{\pi}^t(\Sigma+\epsilon Q) \bigg \vert_{\epsilon=0}\\
&= \prod_{\tau=t}^1 AC_{\pi_\tau}(\phi_{\pi}^\tau(\Sigma)) Q \prod_{\tau=1}^{t}C_{\pi_\tau}(\phi_{\pi}^\tau(\Sigma))^TA^T.
\end{align*}
\end{lemma}

\begin{lemma}
\label{lem:ric_der_prop}
For any $t \!\in\! [0,T]$, $\pi \!\in\!\mathcal{X}^T$, $\Sigma, Q_1, Q_2 \!\in\! S^n_+, a,b \!\in\! \mathbb{R}$
\[
g_\pi^t(\Sigma; aQ_1 + bQ_2) = a g_\pi^t(\Sigma;Q_1) + b g_\pi^t(\Sigma;Q_2)
\]
because a directional derivative is linear in the perturbation. In addition, by operator concavity of the $t$-step Riccati map:
\[
\phi_{\pi}^t(\Sigma + \epsilon Q) \preceq \phi_\pi^t(\Sigma) + \epsilon g_{\pi}^t(\Sigma;Q)
\]
\end{lemma}

\begin{lemma}
\label{lem:tr_ric_der}
For all $t \in [1,T]$, $\pi \in \mathcal{X}^T$, and $Q \in S^n_{+}$, if there exists a constant $\beta < \infty$ such that $\phi_{\pi}^t(\Sigma) \preceq \beta I_{n_y}$, then
\[
\tr\bigl( g_{\pi}^t(\Sigma; Q) \bigr) \leq \beta \eta^t \tr(\Sigma^{-1}Q), \qquad \eta := \beta/(\beta + \uline{\lambda}_W) < 1.
\]
\end{lemma}
\begin{IEEEproof}
We follow the steps of \cite[Thm. 5]{ZAHV_Automatica09} but use the tighter bound $\hat{A}_t \hat{\Sigma}_t \hat{A}_t^T \preceq (\beta - \lambda_W^-) I_n$ in (A.4), which leads to $L^{(l)}_t - L^{(l)}_{t+1} \geq \frac{\lambda_W^-}{\beta} L^{(l)}_t$. Also, since $Q \in S^n_+$ we can decompose it as $Q = \sum_{l=1}^n \lambda_Q^l q_l q_l^T$ and let $\xi^{(l)}(0) = \sqrt{\lambda_Q^l} q_l$.
\end{IEEEproof}
%

\begin{lemma}
\label{lem:som_cont}
Consider two nodes $(x_1,\Sigma)$ and $(x_2, \Sigma)$ with $d_\mathcal{X}(x_1,x_2) \leq \delta$. Let $\Sigma_1$ and $\Sigma_2$ be the updated covariance matrices after applying $u \in \mathcal{U}$ at each node. Then,
\[
\Sigma_1 \succeq \gamma \Sigma_2 + (1-\gamma) W \quad \text{and} \quad \Sigma_2 \succeq  \gamma \Sigma_1 + (1-\gamma) W,
\]
where $0 < \gamma := (1+L_mL_f\delta)^{-1} \leq 1$.
\end{lemma}

\begin{IEEEproof}
Let $M_i := M(f(x_i,u))$ and $\rho_i(\cdot) := \rho_{f(x_i,u)}(\cdot)$ for $i = 1,2$. From Assumption \ref{ass:smm_cont}, $d_\mathcal{X}(f(x_1,u), f(x_2,u)) \leq L_f \delta$ and from Assumption \ref{ass:som_cont}, $M_1 \preceq (1 + L_m L_f \delta) M_2$. Then,
\begin{align*}
\rho_1(\Sigma) &= A(\Sigma^{-1} + M_1)^{-1}A^T + W\\
&\succeq A(\Sigma^{-1} + \gamma^{-1}M_2)^{-1}A^T + W\\
&= \gamma \rho_2(\gamma^{-1}\Sigma) + (1-\gamma) W \succeq \gamma \rho_2(\Sigma) + (1-\gamma) W
\end{align*}
The second result follows analogously.
\end{IEEEproof}

\begin{lemma}
\label{lem:big_lemma}
For $t \in [1,T]$, $\epsilon \geq 0$, $\delta \geq 0$, the reduced tree $\mathcal{T}_t^{\epsilon, \delta}$ contains a set of nodes $\{ (x_t^i, \Sigma_t^i) \mid i = 1, \ldots, K\}$ such that:
\begin{align}
d_\mathcal{X}(x_t^*, x_t^i) \leq &\sum_{\tau = 0}^{t-1} L_f^\tau \delta, \quad \forall i, \label{thm2_x}\\
\Sigma_t^* + \epsilon \biggl( \Gamma_t I_{n_y} & + \sum_{\tau=1}^{t-1} \Gamma_\tau g_{\pi^*(\tau+1)}^{t-\tau}(\Sigma_\tau^*;I_{n_y}) \biggr) \label{thm2_cov}\\
&\succeq \Gamma_t \sum_{i=1}^K \alpha_i \Sigma_t^i + \sum_{\tau=1}^{t-1} \Gamma_\tau (1-\gamma_\tau) \phi_{\pi^*(\tau+1)}^{t-1-\tau}(W) \notag,
\end{align}
where $\pi^* = x_1^*,\ldots,x_T^*$ is the optimal trajectory in \ref{eq:prob2},
\[
\textstyle{0 < \gamma_t := (1+ \sum_{s=1}^t L_f^s L_m \delta)^{-1} \leq 1, \quad\text{and}\quad \Gamma_t := \prod_{s=1}^{t-1} \gamma_s}. 
\]
\end{lemma}

\begin{IEEEproof}[Proof of Lemma \ref{lem:big_lemma}]
We proceed by induction.\\
\textbf{\textit{Base Case}}: If the optimal node $(x_1^*, \Sigma_1^*)$ is discarded, then by Definitions \ref{def:eps_alg_red} and \ref{def:gamma_cross} $T_1^{\epsilon, \delta}$ contains a set of nodes $\{(x_1^k, \Sigma_1^k)\}$ such that $d_\mathcal{X}(x_1^*,x_1^k) \leq \delta$ for all $k$ and $\Sigma_1^* + \epsilon I_{n_y} \succeq \sum_k \alpha_1^k \Sigma_1^k$.\\
\textbf{\textit{Hypothesis}}: Suppose that (\ref{thm2_x}) and (\ref{thm2_cov}) hold for some set $\{(x_t^j, \Sigma_t^j) \mid j = 1,\ldots,J\}$ of nodes in $\mathcal{T}_t^{\epsilon, \delta}$.\\
\textbf{\textit{Induction}}: At time $t+1$, there are sets $\{(x_{t+1}^{ji}, \Sigma_{t+1}^{ji})\}_{i=1}^{K_j}$ in $\mathcal{T}_{t+1}^{\epsilon, \delta}$ corresponding to each node $j$ from time $t$ and satisfying
\begin{equation}
\textstyle{d_\mathcal{X}(x_{t+1}^j, x_{t+1}^{ji}) \leq \delta \text{ and } \Sigma_{t+1}^j \!+\! \epsilon I_{n_y} \!\succeq\! \sum_{i = 1}^{K_j} \alpha_{ji} \Sigma_{t+1}^{ji}}. \label{eq:tree_prune}
\end{equation}
From Lemma \ref{lem:ric_dir_der} for every $\tau = 1, \ldots, t$:
\[
g_{\pi^*(t+1)}^1\biggl(\Sigma^*_t; g_{\pi^*(\tau+1)}^{t-\tau}(\Sigma_\tau^*; I_{n_y})\biggr) = g_{\pi^*(\tau+1)}^{t+1-\tau}(\Sigma_\tau^*;I_{n_y}).
\]
From this and Lemma \ref{lem:ric_der_prop}:
\begin{align*}
\Sigma^*_{t+1} & + \epsilon \sum_{\tau=1}^{t} \Gamma_\tau g_{\pi^*}^{t+1-\tau}(\Sigma_\tau^*;I_{n_y})\\
& = \rho_{\pi^*_{t+1}}(\Sigma^*_t) + \epsilon g_{\pi^*(t+1)}^1\biggl(\Sigma^*_t; \sum_{\tau=1}^{t} \Gamma_\tau g_{\pi^*(\tau+1)}^{t-\tau}(\Sigma_\tau^*; I_{n_y})\biggr)\\
&\succeq \rho_{\pi^*_{t+1}}\biggl(\Sigma_t^* + \epsilon \sum_{\tau=1}^{t-1} \Gamma_\tau g_{\pi^*(\tau+1)}^{t-\tau}(\Sigma_\tau^*; I_{n_y}) + \epsilon\Gamma_t I_{n_y} \biggr)
\end{align*}
Note that $\sum_{\tau=1}^{t-1} \Gamma_\tau(1-\gamma_\tau) + \Gamma_t = 1$. Thus, the terms $(1-\gamma_1), \gamma_1(1-\gamma_2), \ldots, \Gamma_{t-1}(1-\gamma_{t-1}), \Gamma_t$ are the coefficients of a convex combination. Using the hypothesis and Lemma \ref{lem:ric_mon_con}:
\begin{align*}
&\rho_{\pi^*_{t+1}} \biggl(\Sigma_t^* + \epsilon \sum_{\tau=1}^{t-1} \Gamma_\tau g_{\pi^*(\tau+1)}^{t-\tau}(\Sigma_\tau^*; I_{n_y}) + \epsilon\Gamma_t I_{n_y} \biggr)\\
&\succeq \rho_{\pi^*_{t+1}}\biggl( \Gamma_t \sum_{j=1}^J \alpha_j \Sigma_t^j + \sum_{\tau=1}^{t-1} \Gamma_\tau (1-\gamma_\tau) \phi_{\pi^*(\tau+1)}^{t-1-\tau}(W) \biggr)\\
&\succeq \Gamma_t \sum_{j=1}^J \alpha_j \rho_{\pi^*_{t+1}}(\Sigma_t^j) + \sum_{\tau=1}^{t-1} \Gamma_\tau (1-\gamma_\tau) \phi_{\pi^*(\tau+1)}^{t-\tau}(W).
\end{align*}
By hypothesis, $d_\mathcal{X}(x_t^*, x_t^j) \leq \sum_{\tau = 0}^{t-1} L_f^\tau \delta$, and from Lemma \ref{lem:som_cont}:
\[
\rho_{\pi^*_{t+1}}(\Sigma_t^j) \succeq \gamma_t \Sigma_{t+1}^j + (1-\gamma_t) W.
\]
The nodes $\{(x_{t+1}^j, \Sigma_{t+1}^j)\}$ might not be in $\mathcal{T}_{t+1}^{\epsilon, \delta}$ but from (\ref{eq:tree_prune}):
\[
\rho_{\pi^*_{t+1}}(\Sigma_t^j) + \gamma_t\epsilon I_{n_y} \succeq \gamma_t \sum_{i=1}^{K_j} \alpha_{ji} \Sigma_{t+1}^{ji} + (1-\gamma_t) W.
\]
Combining the previous results, we have:
\begin{align*}
&\Sigma^*_{t+1} + \epsilon \sum_{\tau=1}^{t} \Gamma_\tau g_{\pi^*}^{t+1-\tau}(\Sigma_\tau^*;I_{n_y}) + \epsilon \Gamma_{t+1} I_{n_y}\\
&\!\succeq \Gamma_t \!\sum_{j=1}^J \!\alpha_j \biggl(\!\rho_{\pi^*_{t+1}}\!(\Sigma_t^j) \!+ \gamma_t \epsilon I_{n_y} \!\!\biggr) \!+\! \underbrace{\sum_{\tau=1}^{t-1} \Gamma\!_\tau(1\!-\!\gamma_\tau) \phi_{\pi^*\!(\tau\!+1)}^{t-\tau}\!(W)}_{(*)}\\
&\succeq \Gamma_t \sum_{j=1}^J \alpha_j \biggl(\gamma_t \sum_{i=1}^{K_j} \alpha_{ji} \Sigma_{t+1}^{ji} + (1-\gamma_t) W \biggr) + (*)\\
&= \Gamma_{t+1} \sum_{j=1}^J \sum_{i=1}^{K_j} \alpha_j \alpha_{ji} \Sigma_{t+1}^{ji} + \sum_{\tau=1}^{t} \Gamma_\tau (1-\gamma_\tau) \phi_{\pi^*(\tau+1)}^{t-\tau}(W).
\end{align*}
Thus, the set $\bigcup_{j = 1}^J \bigcup_{i = 1}^{K_j} \{ (x_{t+1}^{ji}, \Sigma_{t+1}^{ji})\}$ satisfies (\ref{thm2_cov}) at time $t+1$. It satisfies (\ref{thm2_x}) at $t+1$ from (\ref{eq:tree_prune}) and Assumption \ref{ass:smm_cont}.
\end{IEEEproof}

\begin{IEEEproof}[Proof of Thm \ref{thm:eps_del_red}]
As defined in Lemma \ref{lem:big_lemma} $\Gamma_T = \zeta_T^{-1}$. Define
\[
V(\cdot):= \log\det(\cdot) \text{ and } G := \Gamma_T I_{n_y} + \sum_{\tau=1}^{T-1} \Gamma_\tau g_{\pi^*(\tau+1)}^{T-\tau}(\Sigma_\tau^*;I_{n_y}).
\]
By monotonicity of $V(\cdot)$ and the result in Lemma \ref{lem:big_lemma}:
\[
V(\Sigma^*_T + \epsilon G) \geq V\biggl(\Gamma_T \sum_{i=1}^K \alpha_i \Sigma_T^i + \sum_{\tau=1}^{T-1} \Gamma_\tau (1-\gamma_\tau) \phi_{\pi^*(\tau+1)}^{T-1-\tau}(W) \biggr)
\]
for some set of nodes $\{(x_T^i,\Sigma_T^i) \mid i = 1,\ldots,K\}$ in the reduced tree $\mathcal{T}_T^{\epsilon,\delta}$. By definition, $\phi_\pi^t(W) \succeq W$ for any $t,\pi$ and $\sum_{\tau=1}^{T-1} \Gamma_\tau (1-\gamma_\tau) = 1-\Gamma_T$. By concavity of $V(\cdot)$:
\begin{align}
&V(\Sigma^*_T + \epsilon G) \geq \Gamma_T \sum_{i=1}^K \alpha_i V(\Sigma_T^i) + (1-\Gamma_T) V(W) \label{eq:cor_proof_1}\\
&\geq \Gamma_T V(\Sigma_T^{i^*}) + (1-\Gamma_T) V(W) \geq \Gamma_T V_T^{\epsilon,\delta} + (1-\Gamma_T) V(W). \notag
\end{align}
The second inequality holds because a convex combination of scalars is lower bounded by the smallest one $i^*$. The last inequality holds because $\pi^{\epsilon,\delta}$ is the optimal path in the reduced tree. Next, by concavity of $\log\det(\cdot)$:
\begin{align}
V(&\Sigma^*_T + \epsilon G) \leq V(\Sigma^*_T) + \epsilon \frac{d}{d\epsilon} V\biggl(\Sigma_T^* + \epsilon G \biggr) \bigg \vert_{\epsilon = 0} \notag\\
&= V_T^* + \epsilon \tr\biggl((\Sigma_T^*)^{-1}G\biggr) \leq V_T^* + \epsilon\frac{1}{\uline{\lambda}_W} \tr(G). \label{eq:cor_proof_2}
\end{align}
From Lemma \ref{lem:tr_ric_der} and since $\tr( (\Sigma_T^*)^{-1} ) \leq n_y/\uline{\lambda}_W$:
\begin{align}
\tr(G) &= \Gamma_T \tr(I_{n_y}) + \sum_{\tau=1}^{T=1} \Gamma_\tau \tr\biggl(g_{\pi^*(\tau+1)}^{T-\tau}(\Sigma_\tau^*; I_{n_y}) \biggr) \notag\\
&\leq  n_y \Gamma_T + \sum_{\tau=1}^{T=1} \Gamma_\tau \beta_* \eta_*^{T-\tau} \tr( (\Sigma_T^*)^{-1} ) \leq \Gamma_T \Delta_T \label{eq:cor_proof_3}
\end{align}
Finally, by combining (\ref{eq:cor_proof_1}), (\ref{eq:cor_proof_2}), and (\ref{eq:cor_proof_3}) we get:
\begin{gather*}
\Gamma_T V_T^{\epsilon,\delta} + (1-\Gamma_T)V(W) \leq V_T^* + \epsilon \Gamma_T \Delta_T\\
0 \leq \Gamma_T(V_T^{\epsilon,\delta} - V_T^*) \leq (1-\Gamma_T)(V_T^* - V(W)) + \epsilon \Gamma_T \Delta_T.
\end{gather*}
Multiplying by $\zeta_T = \Gamma_T^{-1}$ gives the result in (\ref{eq:eps_del_guarantee}).
\end{IEEEproof}

\bibliographystyle{IEEEtran}
\bibliography{bib/ICRA14_Bibliography.bib}

\end{document}